\documentclass{article}
\usepackage{graphicx} 
\usepackage{latexsym}
\usepackage{amsmath}
  \usepackage{babel}
  
\usepackage{comment} 
\renewcommand{\epsilon}{\varepsilon}

\usepackage{amsmath,bm}
\usepackage{amssymb}
\usepackage{xcolor}
\usepackage{amsthm}
\usepackage{mathtools}
\usepackage{graphicx}
\usepackage{dsfont}
\usepackage{bbm}
\usepackage{url}
\usepackage{placeins}
\usepackage{multirow}
\usepackage[font=footnotesize,labelfont=bf]{caption}
\usepackage{authblk}
\usepackage{longtable}

\usepackage[authoryear]{natbib}
\usepackage{hyperref}

\def\3{\ss}

\newcommand{\bea}{\begin{eqnarray*}}
	\newcommand{\eea}{\end{eqnarray*}}
\newcommand{\be}{\begin{eqnarray}}
	\newcommand{\ee}{\end{eqnarray}}

\newcommand{\ba}{\begin{array}}
	\newcommand{\ea}{\end{array}}

\def\3{\ss}

\begin{document}
	\title{{\bf \Large Estimating effect thresholds and beyond: A flexible framework for multivariate alert detection
   }}

	\author[1,2]{L. Ameis} 
        \author[1,2,3]{N. Hagemann}
	\author[1,2*]{K. M\"ollenhoff}

	\small\affil[1]{Institute of Medical Statistics and Computational Biology (IMSB), Faculty of Medicine, University of Cologne}
    \small\affil[2]{Division of Mathematics, Department of Mathematics and Computer Science, University of Cologne}
    \small\affil[3]{Department of Biostatistics, Harvard T.H. Chan School of Public Health, Harvard University} \normalsize

	\date{}

	\maketitle

\maketitle

\begin{abstract}
Evaluating 
the influence of 
 continuous covariates, like exposure time or dose, on a response variable is a pivotal objective in the assessment of a compound's effect, particularly when determining toxicity in pre-clinical research or pharmacokinetics in clinical trials. The determination of an alert, such as the $ED_{50}$ value, at which a pre-specified threshold of the response variable is crossed, is an important tool for the evaluation process. In practice, response data might be available for combinations of different covariates 
 and the alert 
depending on both is of interest. In this case, it is crucial to use all available information and extrapolate between cases to ensure the optimal utilization of the data.

In this paper, we introduce a parametric approach that allows alerts to be estimated in a multidimensional setting.  For time-dose-response data, for instance, alert doses at a given time can be determined, even when there are no measurements available at that exact time. 
Likewise, it allows estimation of alert times for a given dose. More generally, the method makes it possible to characterize the complete alert relationship between covariates 
by leveraging all available data. This is achieved by fitting a parametric model and constructing either a confidence band for the two-dimensional curve given for example a fixed time or dose or by constructing a confidence plane for the three-dimensional model fit. Both are derived by a two-step bootstrap approach and summarized in terms of a hypothesis test that can be adjusted to accommodate a variety of alert types. 
The initial model fit is achieved by the flexible framework of  Generalized Additive Models for Location, Scale and Shape (GAMLSS), 
which offers the possibility to account for a plethora of complex three-dimensional data structures.

We demonstrate the validity of our approach through a simulation study and present an application to data from a study investigating the relevance of the exposure duration on cytotoxicity in primary human hepatocytes.

\end{abstract}
\section{Introduction}\label{sec:int}
Many pharmacological and toxicological processes are inherently three-dimensional in nature. They depend not only on the administered dose, but also on the time since exposure or, in combination settings, on the joint action of multiple agents. 
In drug combination studies, this multidimensional relationship can reduce drug resistance and adverse effects while enhancing therapeutic efficacy. This phenomenon has been observed in a variety of research areas, including cancer studies \citep{mokhtari2017combination} and anesthesia research \citep{lee2010drug}.  Similarly, incorporating both exposure time and concentration is essential when assessing compound effects, particularly in the context of determining pharmacokinetics \citep{lange2015analysis} in clinical trials, or analyzing toxicity \citep{Gu18} in pre-clinical research.

When it comes to toxicological data, a recent publication suggests the use of parametric approaches \citep{KD23}. In particular, the identification of alerts in dose-response or time-response data has benefited from parametric modeling approaches (see \cite{MS22,KG21,ameis2024identification}). These methods allow for an interpolation between measured time points or dose levels, thereby improving upon the classical approach of conducting multiple \textit{t}-tests or the recommended Dunnett test \citep{Hot14}. With the recent advancements in  surface modeling for multidimensional  response data, such as in \citet{zhou2025combination} and  \citet{huusari2025predicting} for drug-combinations or \citet{DH22} and \citet{schuettler} for risk assessment, the extension of alert identification to these settings is a natural methodological next step.

A wide range of alert criteria has been proposed. These include the lowest dose at which a noticeable effect is observed \citep{JK19} and the lowest measured dose at which the difference between the dose and the control significantly exceeds  a critical response threshold  \citep{DF11}. Similarly, one could consider the well-known $ED_{50}$ value, or the more general $ED_p$ ($0<p<100$) value \citep{bretz2010practical}, an alert corresponding to the dose, where half (or $p\%$) of the maximum predicted effect is observed. For multidimensional mixture effect modeling, \citet{martin2015defining} introduced a multivariate extension to the  Loewe additivity \citep{loewe1953problem} and,  recently, \citet{Sch25} constructed efficient designs for identifying effective dose combinations. Both approaches are based on  multivariate extensions of the $ED_p$, thereby extending classical alert definitions for three-dimensional response settings.

Alerts that are identified based on multi-dimensional data incorporate more than one covariate, which is often necessary to obtain a comprehensive understanding of the biological background. For instance, considerable variations in the toxicity of compounds can be observed across varying dose levels and exposure durations in cytotoxicity data \citep{bsp12day1,Gu18,riddell1986importance}. The identification of alerts, such as the $ED_{50}$, exclusively for specific exposure periods, has the potential to result in under- or overestimation of the toxic potential. Furthermore, the identification of a multi-dimensional alert based on multiple exposure durations, in conjunction with varying dose levels, facilitates interpolation and the incorporation of all available data. The latter approach can help avoid overfitting when compared with the fitting of curves to individual exposure durations and the subsequent inference of results \citep{DD23}. A second example are drug combinations studies. As previously mentioned, the combination of drugs has been demonstrated to reduce negative and enhance positive effects.  The identification of multi-dimensional alerts, such as the multiple effective dose (MED), as for example investigated in \citep{Sch25}, allows for the quantification of their combined effect. This, in turn, can assist in the assessment of suitable combinations. For example, known reverse effects of specific dose levels of one of the drugs can be avoided.

 In this paper, a model-based approach is proposed that allows the identification of alerts in a three-dimensional data situation, such as, for example, time-dose-response data or the combined effect of two drugs in a drug combination study.
Inspired by \citet{MS22} and \citet{KG21}, this is achieved through the formulation of a hypothesis test, thereby basing alert identification on statistical significance.  The test is formulated for either a two-dimensional question or a three-dimensional question that considers a test statistic 
based on whether the response signal significantly exceeds a pre-specified relevance threshold $\lambda$.
In the  formulation of the hypothesis, the response signal is centered around a reference level, for example, a placebo dose. Depending on the data structure, this might not be necessary and the formulation can be simplified. These simplifications are also presented. 
In all cases, rejecting the null hypothesis indicates an alert, defined as the smallest dose or time point (per time point or dose level, respectively) for which the null hypothesis is rejected. Depending on the precise formulation of the hypothesis, the test decision is based on the construction of a lower or upper simultaneous confidence band for the two-dimensional question or a lower or upper simultaneous confidence plane for the three-dimensional question. These are estimated via an extension of the bootstrap approach introduced in \citet{MS22}.

The proposed method is specifically designed to accommodate complex three-dimensional data structures. Both the two-dimensional and three-dimensional approaches are based on an initial three-dimensional model fit, for which the framework of Generalized Additive Models for Location, Scale and Shape (GAMLSS) \citep{rigby2005generalized} is utilized. This introduces a high degree of flexibility and allows for a wide range of modeling extensions. In contrast to classical mean-based regression approaches, GAMLSS enable the simultaneous modeling of multiple distributional parameters, including variability. This is particularly relevant when considering heteroscedasticity depending, for example, on the dose level or exposure duration in a complex manner.
The GAMLSS approach integrates a multitude of adjustments, such as the inclusion of random effects or spline based modeling, and is continuously developed and well integrated in \texttt{R} \citep{Gamlssside}. Therefore, this allows for the generation of bootstrap samples that capture the intricate characteristics of the original data set.

This paper is structured as follows: Initially, the methodology for identifying alerts and alert relationships is introduced based on an estimated simultaneous confidence band or plane, respectively. 
Subsequently, the approach is evaluated in a simulation study. Finally, the application of the method is demonstrated using a data set from a cytotoxicity study \citep{Gu18} that investigates the relevance of the exposure duration in primary human hepatocytes.

\section{Model fit} \label{sec:model}
Although the proposed approach is generally applicable to any two or even more continuous variables, we deploy a notation based on its most relevant application of time-dose-response data for notational simplicity. 
Fundamentally, the idea of this method is to initially fit a parametric regression model to the data, and subsequently evaluate the response by determining an alert of interest based on a hypothesis test. The procedure was inspired by the work of \citet{MS22} and adapted to the more complex nature of a three-dimensional data structure. In this section, the initial model fit is discussed.

Let the data set under consideration consist of $n$ observations in total, collected at $m_t+1$ different time points, $t_p\in T\subset \mathbb{R}$, where $p = 0, ..., m_t$, and at $m_d+1$ different dose levels, $d_q\in D\subset \mathbb{R}$, where $q = 0, ..., m_d$. Without loss of generality, it is assumed that $t_0$ and $d_0$ are the respective reference or placebo levels. 
Let $n_{p,q}$ denote the number of observations taken at the time-dose combination $(t_p,d_q)\in T\times D$. After a possible preliminary  selection step from a pool of candidate models to ensure goodness of fit - using, for example, the AIC \citep{FK22} as criteria - the initial step for all variations of the proposed approach is to fit a parametric model to the data. 
There are a multitude of  models that adjust for the challenges of complex data structures like autocorrelation or random effects \citep{box2015time,galecki2012linear,demidenko2013mixed}. 
In addition to other applications, this is particularly relevant in time-dose-response models, where residual variability can be heteroscedastic and may depend on explanatory variables.
Therefore, in this manuscript, the flexible approach offered by Generalized Additive Models for Location, Scale, and Shape (GAMLSS) \citep{rigby2005generalized} is chosen. The GAMLSS framework allows all the
parameters of the response variable distribution to be modeled as regression functions of the explanatory variables. 
In (time-)dose-response models, often normality is assumed (see, e.g., \citet{MS22,ameis2024identification}). Hence, we will focus on normally distributed responses, despite the GAMLLS being also applicable to non-normal response variables. 
Let
$$y_{pq}\sim \mathcal{N}(\mu_{pq},\sigma_{pq}^2)$$
with $p=0,\dots,m_t$ and $q=0,\dots,m_d$.
In this case, the GAMLSS framework enables modeling both the mean ($\mu$) and the standard deviation ($\sigma$).  Precisely,  the regression models under consideration are
\begin{align}
\begin{aligned}
    \text{id}(\mu_{pq})&=\mu_{pq}=f(t_p,d_q,\bm{\theta})\\
    \log(\sigma_{pq})&=\eta_{pq}=g(t_p,d_q,\bm{\vartheta})
\end{aligned}
\end{align}
which depend on  time and dose. Hereby, the identity and the $\log$-function serve as link functions.
For an observation $y_{pqr}$ this means
$$y_{pqr}=f(t_p,d_q,\bm\theta)+\epsilon_{pqr},$$
whereby $\epsilon_{pqr}\sim\mathcal{N}(0,\sigma^2_{pq})=\mathcal{N}(0,\exp(g(t_p,d_q,\bm\vartheta))^2)$.\\

GAMLSS allow for different effect types, including linear, non-linear or spline-based models, and random effects \citep{GAMLSSRpaper}. Here, the emphasis is placed  on parametric non-linear regression for the mean  ($f$) and linear regression  for the standard deviation ($g$) for the sake of simplicity.  The estimation of GAMLSS is based on maximizing the penalized likelihood as described in, for example,  \citet{GAMLSSbook}. 
However, as the regression models under consideration, i.e. parametric linear and non-linear models, do not involve penalized terms, the estimation reduces to maximum likelihood estimation \citep{GAMLSSRpaper}.

\section{Alert detection}\label{sec:meth}


After the initial model fit, the method can be applied in two distinct ways.  Firstly, the parametric nature of the approach makes it possible to determine the alert for a fixed dose level or time point, even where no measurements were taken. Secondly, it can be employed to predict the two-dimensional alert relationship between time and dose. In this section, both applications are discussed.

\subsection{Hypothesis test: Two-dimensional question}

In the following, the approach for a fixed time point $\tilde{t}\in T$ is presented, thereby projecting from three-dimensional data to a two-dimensional question. The procedure can be conducted analogously for a fixed dose level. In this two-dimensional situation, the alert corresponds to the minimum dose, where the response (centered on the response at the placebo level) significantly exceeds a pre-specified relevance threshold $\lambda$, with respect to the fixed time point $\tilde{t}\in T$. In terms of a hypothesis test, this can be expressed as follows:

\begin{align}\label{hypotheses}
\begin{aligned}
&H_0:  \forall \ d\in D: \  \big|f(\tilde{t},d,\bm{\theta}))- f(\tilde{t}, d_0,\bm{\theta})\big| \leq\lambda \\\text{ vs. } &H_1:  \exists \ d\in D: \  \big| f(\tilde{t},d,\bm{\theta}))- f(\tilde{t},d_0,\bm{\theta})\big|>\lambda
\end{aligned}
\end{align}

for a fixed $\tilde t\in T$. Rejecting $H_0$ means detecting an alert. 
The test decision is conducted as described in \citet{MS22} by means of the lower simultaneous confidence band to the $\alpha$-level, thereby soliciting a test at the significance level $\alpha$. Precisely,  let $LB^\alpha(\tilde{t},d,\bm{\hat\theta})$ denote the lower simultaneous $(1-\alpha)$-confidence band of $\big|f(\tilde{t},d,\bm{\theta}))- f(\tilde t, d_0,\bm{\theta})\big|$, i.e.
\begin{align}\label{lowercb}
\mathbb{P}\big\{\forall d\in D: LB^\alpha(\tilde t,d,\bm{\hat\theta}) \leq \big|f(\tilde t,d,\bm{\theta}))- f(\tilde t, d_0,\bm{\theta})\big|\big\}\geq 1- \alpha.\end{align}
If
$LB^\alpha(\tilde t,d,\bm{\hat\theta})>\lambda$, $H_0$ is rejected
and the alert dose is defined as the minimum dose, where $H_0$ is rejected, i.e.
$\min\big\{d\in D : LB^\alpha(\tilde t,d,\bm{\hat\theta})>\lambda\big\}.$
This procedure results in a test at the $\alpha$-level \citep{MS22}. 

\begin{figure}[h]
    \centering
    \includegraphics[width=0.6\linewidth]{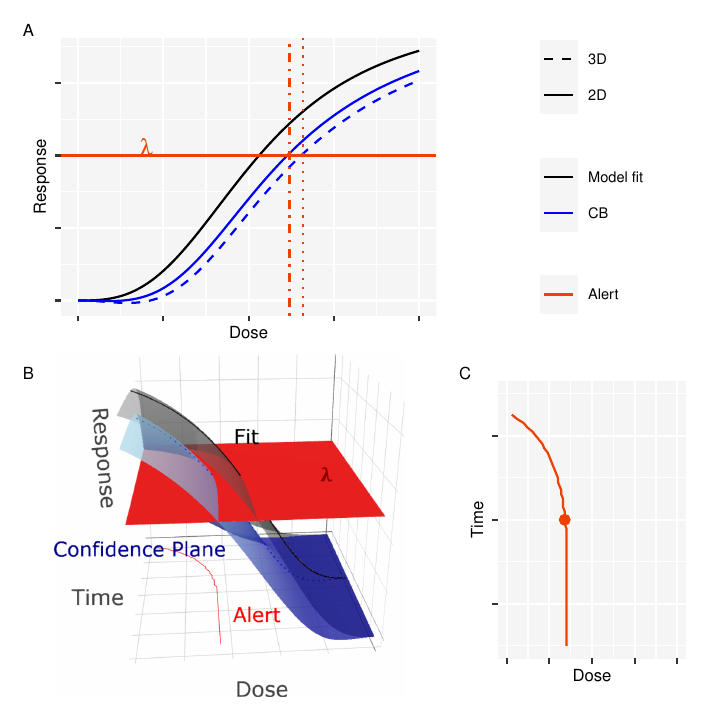}
    \caption{ (A) Centered absolute value of the model fit and estimated confidence bands for a fixed time point. The solid blue line indicates the confidence band estimated from the two-dimensional algorithm.  The dashed line indicates a confidence band inferred from the three-dimensional analysis. The red horizontal line marks the threshold value. The red vertical lines visualize the estimated alert doses.  (B) Visualization of the three-dimensional approach. The centered absolute values of the model fit are indicated by the black plane. The estimated confidence plane is displayed in blue. The threshold $\lambda$ is indicated by the red hyperplane orthogonal to the $Response$-axis. The time-dose-alert relationship is projected onto the hyperplane spanned by the time-axis and the dose-axis. The black and blue lines exemplify a possible fixed time point. (C) Estimated time-dose-alert relationship resulting from the three-dimensional approach. That is the intersection between the blue surface and the red plain. The dot indicates the alert for the fixed time point.}
    \label{fig:exp}
\end{figure}

 Figure \ref{fig:exp} (A) depicts a visualization where the black and blue lines represent $f$ and the confidence band, respectively. The solid blue line indicates the confidence band estimated from the two-dimensional approach, while the dashed line indicates the confidence band inferred from the three-dimensional algorithm described in Section \ref{sec:met:3d}. The red horizontal line shows the threshold value $\lambda$ and the red vertical line shows the identified alert.

\textcolor{black}{The centering employed in the formulation of the approach ensures that the expression $|f(\tilde t, d, \bm{\theta}) - f(\tilde t, d_0, \bm{\theta})|$ is $0$ at the reference level,
which
 facilitates a streamlined selection of the relevance threshold $\lambda$. 
 Additionally, 
 $|f(\tilde t, d, \bm{\theta}) - f(\tilde t, d_0, \bm{\theta})|$ increases from $0$ at $d_0$. Hence,the threshold value is positive and it is tested whether the threshold is exceeded.
Defining the alert dose as the minimum dose at which the threshold is exceeded ensures the uniqueness of the identified dose level, even if the increase is non-monotonic.}

\textcolor{black}{Depending on the data, the approach can be simplified, when centering to the reference level adds an unnecessary extra step.  
An illustrative example is cell viability data, as presented in the case study in Section \ref{sec:case}. 
Here, the response values are measured in percentage, start at approximately $100\%$ cell viability and tend to $0\%$. An alert of interest could be, when the cell viability falls below a value of $p\%$, e.g. $50\%$, and the relevance threshold would correspondingly be chosen  as $p\%$. Centering would result in values of the form $|f(\tilde t, d, \bm{\theta}) - 100|$, which introduces unnecessary complexity into the interpretation. 
In such cases, the formulation of the hypothesis can be based on $f(\tilde t,d,\bm\theta)$ instead of $|f(\tilde t, d, \bm{\theta}) - f(\tilde t, d_0, \bm{\theta})|$. However,  it then depends on whether the interest lies in $f(\tilde t,d,\bm\theta)$ undercutting or exceeding the pre-specified threshold for any $d\in D$. That is}


\begin{align}
\begin{aligned}\label{hypothesesdecreasing}
\text{Threshold}&\text{ undercut:}\\
&H_0:  \ \forall \ d\in D: \ f(\tilde t,d,\bm{\theta})) \geq\lambda  &\text{   vs. } &H_1: \ \exists \ d\in D: \ f(\tilde t,d,\bm{\theta}))<\lambda\\
\end{aligned}\\
\begin{aligned}\label{hypothesesincreasing}
\text{Threshold}&\text{ exceeded:}\\
&H_0:  \ \forall \ d\in D: \ f(\tilde t,d,\bm{\theta})) \leq\lambda &\text{   vs. } &H_1:\ \exists \ d\in D: \ f(\tilde t,d,\bm{\theta}))>\lambda
\end{aligned}
\end{align}
for a fixed $\tilde{t}\in T$.
In case \eqref{hypothesesincreasing}, the test decision is made using the lower simultaneous confidence band of $f(\tilde t,d,\bm\theta)$ as described above. In case \eqref{hypothesesdecreasing}, the test is instead conducted using an upper simultaneous confidence band $UB^\alpha(\tilde t,d,\bm{\hat\theta}) $ of $f(\tilde t,d,\bm\theta)$, which is defined analogously to \eqref{lowercb} as:

\begin{align}\label{uppercb}
\mathbb{P}\big\{\forall d\in D: UB^\alpha(\tilde t,d,\bm{\hat\theta}) \geq f(\tilde t,d,\bm{\theta}))\big\}\geq 1- \alpha.\end{align}

 Analogously to the test decision based on the lower simultaneous confidence band, $H_0$ is rejected if $ UB^\alpha(\tilde t,d,\bm{\hat\theta})<\lambda$
and the alert dose is defined as
$\min\big\{d\in D : UB^\alpha(\tilde t,d,\bm{\hat\theta})<\lambda\big\}.$
 \textcolor{black}{As explained above, taking the minimum ensures the unique identification of an alert dose, even if $f$ is non-monotonic. }
 
\subsection{Hypothesis test: Three-dimensional approach} \label{sec:met:3d}

The three-dimensional approach is similar overall, but instead of focusing on results for a fixed time point or dose level, it aims to identify the  time-dose-alert relationship. 
Let $(t_0, d_0)\in T\times D$ be the combination of the reference and the placebo level. In this three-dimensional framework, the hypothesis test can be expressed as follows:

\begin{align}\label{hypotheses3dim}
\begin{aligned}
&H_0: \forall \ (t,d)\in T\times D: \  \big|f(t,d,\bm{\theta})- f( t_0, d_0,\bm{\theta})\big| \leq\lambda \\\text{ vs. } &H_1: \exists \ (t,d)\in T\times D: \  \big| f(t,d,\bm{\theta})- f( t_0, d_0,\bm{\theta})\big|>\lambda.
\end{aligned}
\end{align}

Rejecting $H_0$ again means detecting an alert.
Rather than using a confidence band for a two-dimensional curve, the test decision is reached via a  lower simultaneous confidence plane $LP^\alpha\big(t,d, \bm{\hat\theta}\big)$  of $\big|f(t,d,\bm{\theta}))- f( t_0, d_0,\bm{\theta})\big|$ to the $\alpha$-level.  This confidence plane fulfills  
$$\mathbb{P}\Big\{\forall (t,d)\in T\times D: LP^\alpha\big(t,d,\bm{\hat\theta}\big) \leq \big|f(t,d,\bm{\theta}))- f( t_0, d_0,\bm{\theta})\big|\Big\}\geq 1- \alpha.$$

Conceptually, the idea is the same as above, but defined for the whole plane. Again, $H_0$ in \eqref{hypotheses3dim} is rejected if $LP^\alpha\big(t,d, \bm{\hat\theta}\big)>\lambda.$ The procedure results in a test to the $\alpha$-level since
\begin{align*}
\mathbb{P}_{H_{0}}\{\exists (t,d)\in T\times D: LP^\alpha(t,d,\bm{\hat{\theta}}) >\lambda \}
&\leq\mathbb{P}_{H_{0}}\{\exists (t,d)\in T\times D: LP^\alpha(t,d,\bm{\hat{\theta}}) >\big|f(\tilde{t},d,\bm{\theta}))- f(\tilde{t}, d_0,\bm{\theta})\big|\}\\
&=1-\mathbb{P}_{H_{0}}\{\forall (t,d)\in T\times D: LP^\alpha(t,d,\bm{\hat{\theta}}) \leq\big|f(\tilde{t},d,\bm{\theta}))- f(\tilde{t}, d_0,\bm{\theta})\big| \}\\
&\leq1-(1-\alpha)=\alpha.
\end{align*}
A visualization is depicted in Figure \ref{fig:exp} (B). Here, the detected alert is not a single dose level indicated by a vertical red line, but rather the time-dose-alert relationship, projected onto the hyperplane spanned by the time-axis and the dose-axis. Additionally, this projection is depicted in Figure \ref{fig:exp} (C).



This theory naturally yields the alert for a fixed time point (or dose level) as a special case, which can be determined directly from the estimated confidence plane. The two-dimensional projection of the resulting confidence band is illustrated as the blue dashed line in Figure \ref{fig:exp} (A). By nature of construction (see Section \ref{sec:meth:cb}), the resulting confidence band is wider than that estimated using the two-dimensional approach.
Simplified hypotheses are analogous to \eqref{hypothesesdecreasing} and \eqref{hypothesesincreasing}.

\subsection{Testing procedure}\label{sec:meth:cb}
In principle, the test can be conducted using any method of constructing a lower or upper simultaneous confidence band for a two-dimensional curve, or a lower or upper simultaneous confidence plane in three dimensions, respectively. This manuscript utilizes and extends the procedure introduced in  \citet{MS22} due to its natural flexibility.
Let $f(t,d,\bm{\hat{\theta}})$ be the estimated parametric model fit of the mean. Let
$$\Delta(t,d,\bm{\hat\theta})=\begin{cases}
      \big|f(\tilde t,d,\bm{\hat\theta}))- f( \tilde t, d_0,\bm{\hat\theta})\big|, & \text{two-dimensional case, centered} \\
    f(\tilde t,d,\bm{\hat\theta}),& \text{two-dimensional case, simplified}\\
    \big|f(t,d,\bm{\hat\theta}))- f( t_0, d_0,\bm{\hat\theta})\big|, & \text{three-dimensional case, centered} \\
    f(t,d,\bm{\hat\theta}),& \text{three-dimensional case, simplified}
\end{cases}$$
 depending on whether the two- or three-dimensional hypothesis is regarded and whether it is centered or simplified.
 The lower simultaneous confidence plane is constructed based on an estimator for the standard deviation of $\Delta(t,d,\bm{\hat\theta})$ and a quantile $c$, such that 

$$\mathbb{P}\Bigg\{\max_{(t,d)\in T\times D}\frac{\Delta\big(t,d,\bm{\hat\theta}\big)-\Delta\big(t,d,\bm{\theta}\big)}{\hat{\sigma}_{\Delta\big(t,d,\bm{\hat{\theta}}\big)}}\leq c_{lower,3dim}\Bigg\}=1-\alpha$$
in the three-dimensional case.
For the two-dimensional case, the maximum is taken over the subset $ \{\tilde t\}\times D$. 
Hence, the lower simultaneous confidence band or plane can be constructed as
\begin{align}
   L^\alpha(t,d,\bm{\hat{\theta}}) \coloneqq \Delta(t,d,\bm{\hat\theta})-c_{lower,2dim/3dim}\hat{\sigma}_{\Delta(t,d,\bm{\hat\theta})},
\end{align}
whereby $L$ denotes either $LB$ or $LP$.
It holds $LB^\alpha(\tilde t,d,\bm{\hat\theta})\geq LP^\alpha(\tilde t,d,\bm{\hat\theta})$, since $c_{lower,2dim}\leq c_{lower,3dim}$, as the maximum difference between the true and estimated value is taken over a subset in the two-dimensional case. 
Naturally, this can also be explained because the curve for a fixed time point $\tilde t$ is part of the  three-dimensional plane, and a simultaneous confidence plane to the $\alpha$-level also encapsulates this curve to the $\alpha$-level. However, if only a fixed time point is of interest, the approach based on the confidence plane will inherently yield a more conservative test.

In order to estimate $LB^\alpha(\tilde t,d,\bm{\hat{\theta}})$ or $LP^\alpha(t,d,\bm{\hat{\theta}})$, it is necessary  to estimate $c_{lower,3dim/2dim}$ and $\hat\sigma_{\Delta(t,d,\bm{\hat\theta})}$. 
This is achieved via a nested two-level bootstrap procedure as introduced in \cite{MS22}. This approach resembles the classic bootstrap-t intervals introduced in \citet{ET94}, but is applied to the entire plane or curve. Calculating the maximum for the estimation of $c$ ensures that $LB^\alpha(\tilde t,d,\bm{\hat{\theta}})$ or $LP^\alpha(t,d,\bm{\hat{\theta}})$ are blown up from enclosing the curve or plane pointwise to simultaneously enclosing it at $\alpha$-level.Precisely, $B_1$ bootstrap samples are generated similar to the original data set (for details see Section \ref{sec:meth:est}). The empirical standard deviation $\hat\sigma_{\Delta(t,d,\bm{\hat\theta})}$ is then estimated from $\Delta(t,d,\bm{\hat\theta_1^*}),\dots,\Delta(t,d,\bm{\hat\theta_{B_1}^*})$. Additionally, for each generated first-level bootstrap sample $l=1,\dots,B_1$, a second level of $B_2$ bootstrap samples is generated. The sample generation step is analogous, but the first-level bootstrap sample is used instead of the original data set as a basis. From each set of $B_2$ second-level bootstrap samples, the empirical standard deviation  
$\hat\sigma_{\Delta(t,d,\bm{\hat\theta_l^*})}$ is estimated. These are then used to calculate

$$D^{*,l}=\max_{(t,d)\in T\times D}\frac{\Delta\big(t,d,\bm{\hat\theta_l^*}\big)-\Delta\big(t,d,\bm{\hat\theta}\big)}{\hat{\sigma}_{\Delta\big(t,d,\bm{\hat{\theta}_l^*}\big)}},$$ from which $c$ is estimated as the empirical $(1-\alpha)$-quantile. 

Sample generation is identical for the two- and three-dimensional approaches. The other steps are similar, with the only difference being that $t=\tilde t$ is fixed for the two-dimensional case.
The upper simultaneous confidence band or plane, $UB^\alpha(\tilde{t},d,\bm{\hat{\theta}})$ or $UP^\alpha(t,d,\bm{\hat{\theta}})$, is constructed and estimated analogously.

\subsection{Generation of bootstrap samples and faster algorithm}\label{sec:meth:est}
There is a fundamental interconnection between the model assumption, estimation, and bootstrap sample generation. 
As discussed in Section \ref{sec:model}, fitting GAMLSS to the data results in estimators $\bm{\hat\theta}$  and $\bm{\hat\vartheta}$ for the regression models of the mean $\mu$ and standard deviation $\sigma$, respectively. Accordingly, an observation $y^*_{pqr}$ at $(t,d)\in T\times D$ for a first-level bootstrap sample is generated by drawing  errors $\epsilon^*_{pqr}\sim\mathcal{N}\big(0,\exp(g(t,d,\bm{\hat\vartheta})^2\big)$ and calculating
$$y^*_{pqr}=f(t,d,\bm{\hat\theta})+\epsilon^*_{pqr}.$$
Let $\bm{\hat\theta}^*_l$  and $\bm{\hat\vartheta}^*_l$, $l=1,\dots,B_1$, denote the estimated parameters from a first-level bootstrap sample. An observation of a second-level bootstrap is generated similarly by drawing errors $\epsilon^{**}_{pqr}\sim\mathcal{N}\big(0,\exp(g(t,d,\bm{\hat\vartheta}^*_l)^2\big)$ and calculating
$$y^{**}_{pqr}=f(t,d,\bm{\hat\theta}^*_l)+\epsilon^{**}_{pqr}.$$

The estimation procedure for  GAMLSS requires more  computational effort than a simple OLS or maximum likelihood (ML) estimation. Therefore, reducing the computation time might be of interest if the approach is applied to a large number of data sets. Due to the model assumptions, both $\bm{\hat\theta}^{**}$ and $\bm{\hat\vartheta}^{**}$ are estimated for each of the second-level bootstrap samples, respectively. However, only  $\bm{\hat\theta}^{**}$ is utilized further in the algorithm.
Therefore, it is possible to simplify the estimation in this step and use OLS/ML to estimate only $\bm{\hat\theta}^{**}$ assuming  iid. error terms $\epsilon_{pqr}\sim N(0,\sigma ^2)$.
This more simplistic assumption has the potential to compromise the precision of the model fit, yet  reduces the time required for computation.

In general, both algorithms produce similar results. Due to the sake of brevity, the results obtained by applying the fast algorithm are presented in this manuscript. The results of the normal algorithm are presented in the supplementary material. Note that both algorithms are identical  if a constant standard deviation is assumed.


\section{Simulation Study}\label{sec:sim}
\subsection{Setting and data generation}\label{sec:simscen}
In order to evaluate the proposed approach, a simulation study is conducted. To get a broad view of the method's performance, two distinct scenarios are considered, each based on a different data type. Scenario 1 is based on the time-dose-response  study analyzing cytotoxicity presented as a case study in Section \ref{sec:case}. The response variable is cell viability, measured as a percentage, ranging from around $100\%$ to $0\%$. Measurements were taken at $3$ time points: day $t_0=1$, day $t_1=2$, and day $t_2=7$. For each time point, $6$ equidistant dose levels on a $\log$ scale with base $=\sqrt{10}$ were measured. The alert of interest is $50\%$ cell viability. Therefore, the simplified assumption for undercutting a threshold as presented in \eqref{hypothesesdecreasing} applies and an upper simultaneous confidence band/plane is estimated.  

For the parametric model, 
the \textit{td2pLL} model introduced in \citet{DH22} is selected. See the supplementary material for details on the parametrization.  
To evaluate the  benefits of the GAMLSS framework,  four different cases are compared: constant standard deviation (simple) and  a standard deviation represented by the linear model
\begin{align}\label{sigmaform}
     \log(\sigma) = \beta_{Intercept} +\beta_{Dose}Dose + \beta_{Dose^2}Dose^2 + \beta_{Time}Time + \beta_{Dose^2\cdot Time}Dose^2\cdot Time 
\end{align}
with small, medium and large coefficients. The formulation of the linear model is motivated by the findings of \citet{DH22}. 
The parameters used for the simulation are obtained from Case 2, as presented in Section \ref{sec:case}.
Two sample sizes, $n=250$ and $n=54$, are considered. 

All cases are analyzed applying two distinct assumptions: one assuming a constant standard deviation, and the other assuming the more complex linear dependency shown above. Both the three-dimensional and two-dimensional approaches are utilized. 
For the two-dimensional approach, the fixed time point of $t = 4$ is selected. This choice is consistent with the fixed time point selected in the case study in Section \ref{sec:case}.

Scenario 2 is based on the simulation study presented in \citet{Sch25}, where the optimal design for identifying effective dose combinations within the context of drug combination studies was investigated. The example illustrates  a clear positive interaction between two drugs, here denoted Dose 1 and Dose 2 for simplicity, and an  overall increase in response. Precisely, this means that the response of the individual drugs is increasing, and that the combination of these drugs produces a response that exceeds the maximum response of the individual drugs. The underlying parametric model is a drug-combination  model, whereby both compounds follow the form of an \textit{Emax} model \citep{bornkamp2009mcpmod} (for details see  \citet{Sch25} or the supplementary material). The alerts of interest are $p\%$ multivariate effective dose (MED) contour lines defined as
$$\text{MED}_{p}\big(\bm{\theta}\big)=\Bigg\{(d_1,d_2)\in D_1\times D_2: \frac{f(d_1,d_2,\bm{\theta})-\min_{(d_{1,0},d_{2,0})\in D_1\times D_2}f(d_{1,0},d_{2,0},\bm{\theta})}{R_{\max}}=\frac{p}{100}\Bigg\}$$
with $R_{\max}=\max_{(d_1,d_2)\in D_1\times D_2}f(d_1,d_2,\bm{\theta})-\min_{(d_1,d_2)\in D_1\times D_2}f(d_1,d_2,\bm{\theta})$. 
This concept is a two-dimensional extension of the $ED_{p}$ value, whereby the predicted effect is calculated from the minimum and maximum over all dose pairs instead of from the lower and upper asymptotes.
Since the alerts can be interpreted without centreing, the simplified assumption described in \eqref{hypothesesincreasing} can be applied
and a lower simultaneous confidence band/plane is fitted. The scenario is analyzed using both the three-dimensional and two-dimensional approaches.
 
\begin{figure}[h]
    \centering
    \includegraphics[width=0.75\linewidth]{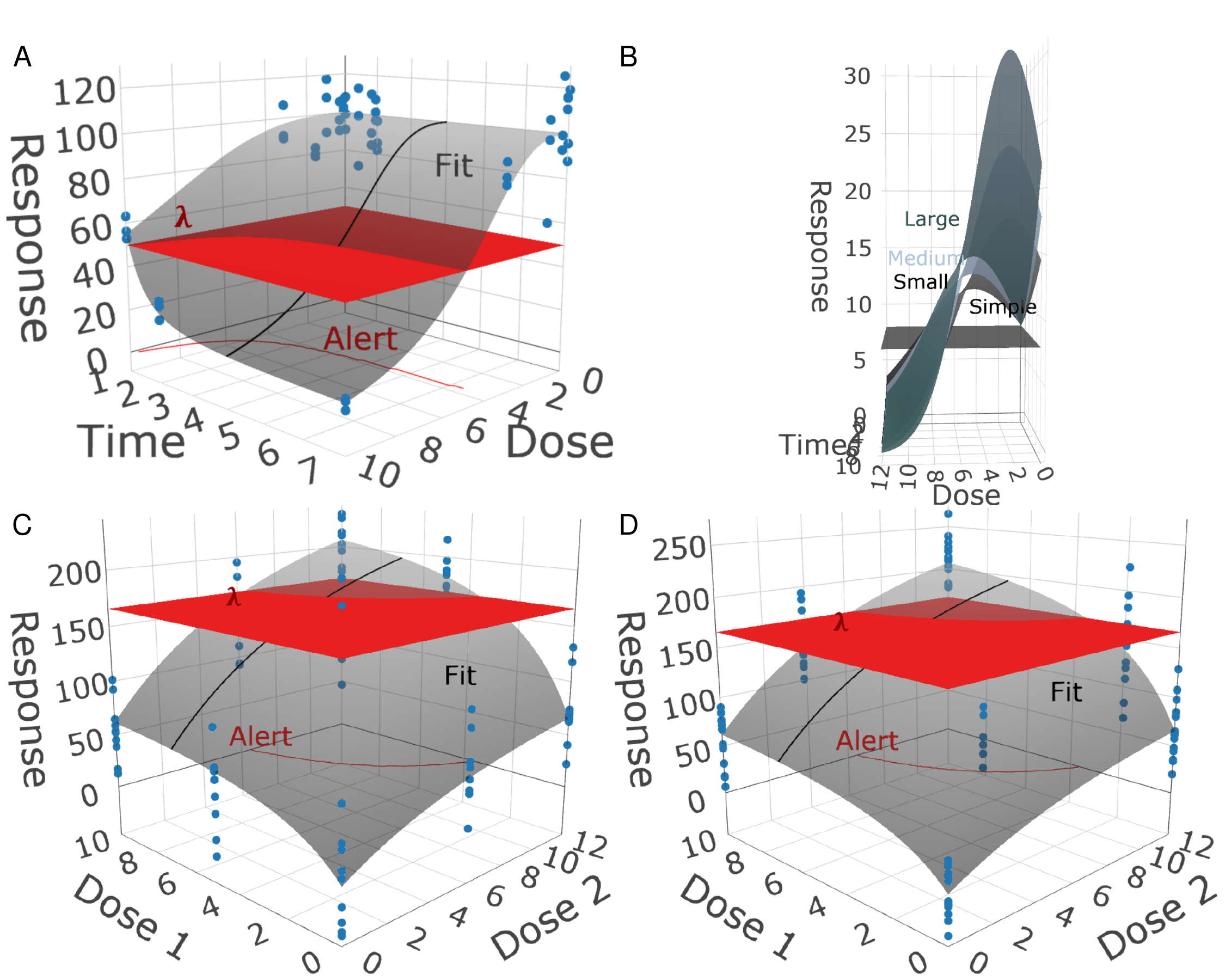}
            \caption{(A) True mean response of Scenario 1. (B) Comparison of the simple, small, medium and large linear models used to generate the standard deviation. (C) Scenario 2 with the factorial $3\times3$ design. (D) Scenario 2 with the D-optimal design. The black plane in (A), (B) and (C) indicate the true response planes. The black lines indicate the two-dimensional curve for the fixed time point $\tilde t=4$ or dose level of Dose 1 $\tilde d_1=7$, respectively. The red plane displays the respective thresholds. The intersection between the threshold plane and the true response marks the true alert relationship between time and dose or dose and dose. For better visibility, this relationship is projected to the $0$-response axis. The blue dots indicate an example of simulated measurements to demonstrate the desing.}
                    \label{fig:Simscenalt}
\end{figure}

To evaluate the influence of the design, two different designs as presented in \citet{Sch25} are regarded. Precisely, the factorial $3\times 3$ design and the D-optimal design for $(80,90)\%$ MED contour lines as identified in the manuscript are selected. Accordingly, the $80\%$ MED contour line is chosen as the alert of interest. 
The exact dose levels for both drugs for which measurements were taken depend on the design. However, in both cases the dose levels range from $d_{1,0}=0$ to $d_{1,m_{D_1}}=10$ for Dose 1 and from $d_{2,0}=0$ to $d_{2,m_{D_2}}=12$ for Dose 2. For the two-dimensional approach, we choose to fix the dose level of Dose 1 at $\tilde d_{1}=7$ for the demonstration, since no measurements were taken for this dose in either design scheme.
The sample sizes $n=152$, $n=90$ and $n=45$ are considered.
Both scenarios are depicted in Figure \ref{fig:Simscenalt}. For details on the precise parameters, see Table \ref{tab:Simparam}. 
\begin{table}[h]
    \centering
        \caption{Parameters of the simulation scenarios. For details on the design see Table A in the supplementary material. Note that a link function is applied  for the standard deviation as described in Section \ref{sec:meth:est}. For the model parameterizations we refer to Section 1 in the supplementary material.}
    \begin{tabular}{c|c|c|c|c}
    Scenario&Parameters $\bm\theta$&Parameters $\bm\vartheta$&Observations $n$&Design \\
    \hline

    1 - Full - Simple & 2.9,5.9,1.8,4.8 & 2.081 & 250   & Full \\
    1 - Full - Small & 2.9,5.9,1.8,4.8 & 2.081,0.162,-0.019,0.06,-0.001 & 250   & Full \\
    1 - Full - Medium & 2.9,5.9,1.8,4.8 & 2.081,0.215,-0.025,0.08,-0.002 & 250   & Full \\
    1 - Full - Large & 2.9,5.9,1.8,4.8 & 2.081,0.269,-0.032,0.1,-0.002 & 250   & Full \\
    1 - Reduced - Simple & 2.9,5.9,1.8,4.8 & 2.081 & 54    & Reduced \\
    1 - Reduced - Small & 2.9,5.9,1.8,4.8 & 2.081,0.162,-0.019,0.06,-0.001 & 54    & Reduced \\
    1 - Reduced - Medium & 2.9,5.9,1.8,4.8 & 2.081,0.215,-0.025,0.08,-0.002 & 54    & Reduced \\
    1 - Reduced - Large & 2.9,5.9,1.8,4.8 & 2.081,0.269,-0.032,0.1,-0.002 & 54    & Reduced\\
    2 - Factorial 3x3 - N152 & 0,80,3,120,10,0.02 & 3.401 & 152   & Factorial 3x3 \\
    2 - Factorial 3x3 - N90 & 0,80,3,120,10,0.02 & 3.401 & 90    & Factorial 3x3 \\
    2 - Factorial 3x3 - N45 & 0,80,3,120,10,0.02 & 3.401 & 45    & Factorial 3x3 \\
    2 - D-optimal - N152 & 0,80,3,120,10,0.02 & 3.401 & 152   & D-optimal \\
    2 - D-optimal - N90 & 0,80,3,120,10,0.02 & 3.401 & 90    & D-optimal \\
    2 - D-optimal - N45 & 0,80,3,120,10,0.02 & 3.401 & 45    & D-optimal 
    \end{tabular}
    \label{tab:Simparam}
\end{table}

The simulated datasets are generated as described in Section \ref{sec:meth:est}. For each combination of scenario, specification, and number of observations, a total of $1,000$ simulation runs are conducted.
For the proposed approach, a large value ($B_1=500$) is chosen for the estimation of the quantiles in the first-level bootstrap, and a
 small value ($B_2=25$) is chosen for the estimation of the standard error in the second-level bootstrap. Thus, the number of generated bootstrap samples overall is $500\times25=12500$. As recommended in \citet{ET94} and utilized satisfactorily in \citet{MS22} and \citet{ameis2024identification}
, these choices are large enough to ensure precise estimations, yet small enough to avoid excessive computational effort.

 Outcomes of the two-dimensional approach are the number of rejections of $H_0$, which corresponds to the number of detected alerts, and the median and standard deviation of the identified alert dose (of Dose 2). For the three-dimensional approach, one outcome is also the number of rejections of $H_0$. Additionally, the estimated time-dose-alert relationships is examined. 
  Specifically, two main characteristics are examined:
 First, the accuracy of the time period or dose levels of Dose 1 for which alerts were identified. Second, the accuracy of the identified alerts. 
 
Let $\tilde{T}_{true}$ denote the true time period or span of dose levels of Dose 1 for which alerts exist. In Scenario 1, $\tilde{T}_{true}=\left[1.1,7\right]$. The true response does not fall below the threshold value only at $t=1$. In Scenario 2, $\tilde{T}_{true}=\left[3.6,10\right]$. The true response does not exceed the threshold value for dose levels of Dose 1 in the interval $\left[0,3.5\right]$. However, the time period or span of dose levels of Dose 1 estimated from the simulation runs, denoted $\tilde{T}_{est}$, can differ from these intervals. The accuracy of the estimation is measured in  recall, precision, onset error and offset error.
 Precisely:
 \begin{align*}
     \text{Recall}_{est}&= \frac{|\tilde{T}_{true}\cap\tilde{T}_{est}|}{|\tilde{T}_{true}|}   \\
          \text{Precision}_{est}&= \frac{|\tilde{T}_{true}\cap\tilde{T}_{est}|}{|\tilde{T}_{est}|}   \\
          \text{Onset error}_{est}&=\min(\tilde{T}_{est})-\min(\tilde{T}_{true})\\
               \text{Offset error}_{est}&=\max(\tilde{T}_{est})-\max(\tilde{T}_{true})
 \end{align*}
 Recall measures how much of the truth the estimates cover, while precision measures how correct the estimates are \citep{fawcett2006introduction,manning2008introduction}. Hereby, $\tilde T_{true}$ and $\tilde T_{est}$ are discretized by superimposed grid and both values are calculated from the respective cardinalities.  These values are examined primarily to identify potential bulks in the estimated confidence planes that could result in the estimated $\tilde{T}_{est}$   not being connected. The onset error and offset error measure the accuracy of the earliest and latest  estimated time points or smallest and largest estimated dose levels of Dose 1 for which alerts are detected, respectively. 

The accuracy of the estimated alerts is evaluated by means of the root mean squared error (RMSE) for the overlap $\tilde{T}_{true}\cap\tilde{T}_{est}$. Precisely, let $d_{est}(t)$ be the identified alert for $t\in\tilde{T}_{true}\cap\tilde{T}_{est}$ and $d_{true}(t)$ the truth. Then:
 $$\text{RMSE}_{est}=\sqrt{\frac{1}{|\tilde{T}_{true}\cap\tilde{T}_{est}|}\sum_{t\in \tilde{T}_{true}\cap\tilde{T}_{est}} (d_{est}(t)-d_{true}(t))^2}$$
  It is important to note that in the context of a simulation run, the initial model fit may not converge. This was observed on a single occasion for Scenario 1 with the reduced design and medium simulated standard deviation. The run was excluded from further analysis.
 All simulation scenarios were run on an HPC-cluster employing \texttt{R} version 4.4.1.

\subsection{Results of the two-dimensional hypothesis}\label{sec:simres2d}
In this section, the results of the alerts at the fixed time point of $\tilde t = 4$ for Scenario 1, and at the fixed dose level of $\tilde d_{1} = 7$ for Scenario 2, as estimated by the two-dimensional confidence band or the three-dimensional confidence plane, are presented. For Scenario 1, only the results of the fast algorithm are discussed here. A comparison to the normal algorithm is provided in the supplementary material. Overall, the results are similar. There is no such differentiation for Scenario 2.

\begin{figure}[h]
    \centering
    \includegraphics[width=\linewidth]{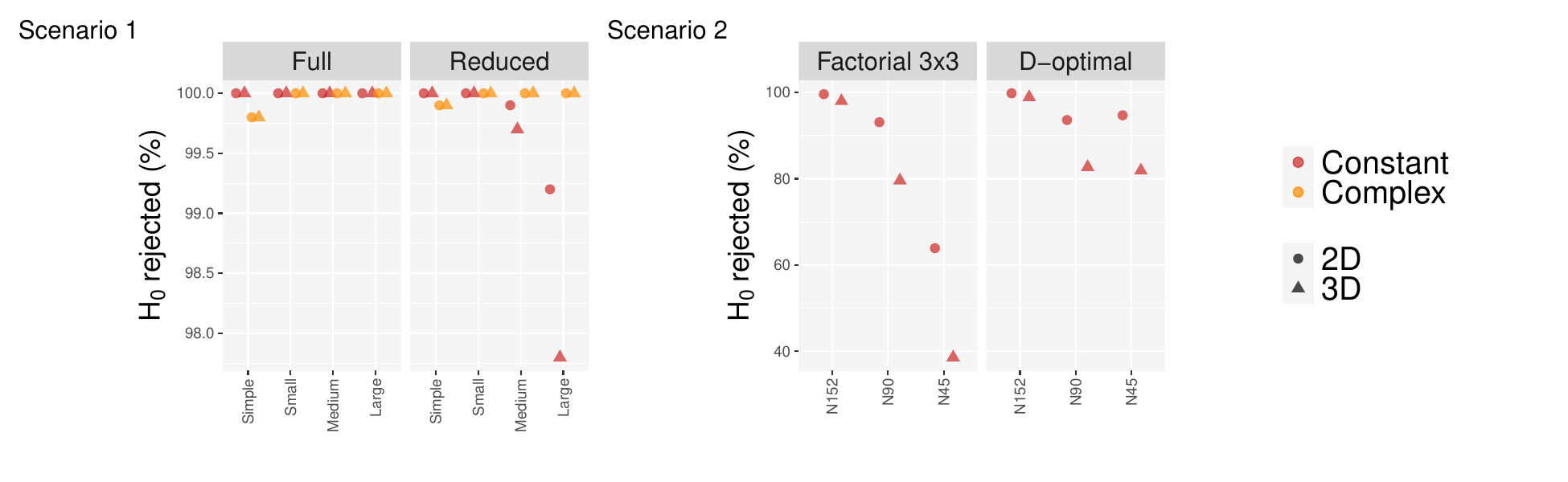}
    \caption{Proportion of rejections of $H_0$ in both scenarios. 
    }
    \label{fig:h0rej}
\end{figure}
Figure \ref{fig:h0rej} and Figure A in the supplementary material, as well as Tables B-E in the supplementary material show the proportion of rejections of $H_0$ in $1,000$ runs, which corresponds to the proportion of detected alerts. As expected, the number of rejections by the two-dimensional approach is greater than or equal to that by the three-dimensional approach, since $c_{3dim}\geq c_{2dim}$. For Scenario 1 with the full design assuming a constant standard deviation, $H_0$ is always rejected. When the complex structure of the standard deviation is assumed correctly, $H_0$ is also always rejected, whereas misspecification in the simple case leads to rejection in at least $99.8\%$ runs. For the reduced design, the complex assumption results in rejections in at least $99.9\%$ runs. For the simple and small cases, $H_0$ is always rejected when assuming a constant standard deviation. For the medium and large cases, however, the number of rejections is slightly smaller, but never dropping below $97.8\%$.

For Scenario 2 with the factorial $3\times3$  design a clear trend of fewer rejections  is observed as the total number of observations decreases, dropping to $63.9\%$ for the two-dimensional approach and $38.6\%$ for the three-dimensional approach. Interestingly, this trend is not observed for the D-optimal design. Although the number of rejections is higher for $n=152$ observations, it remains almost constant for $n=90$ and $n=45$ observations. Additionally, the overall number of rejections is higher for the  D-optimal design, exceeding at least $93.6\%$ for the two-dimensional approach and $81.9\%$ for the three-dimensional approach.

\begin{figure}[h!]
    \centering
    \includegraphics[width=0.8\linewidth]{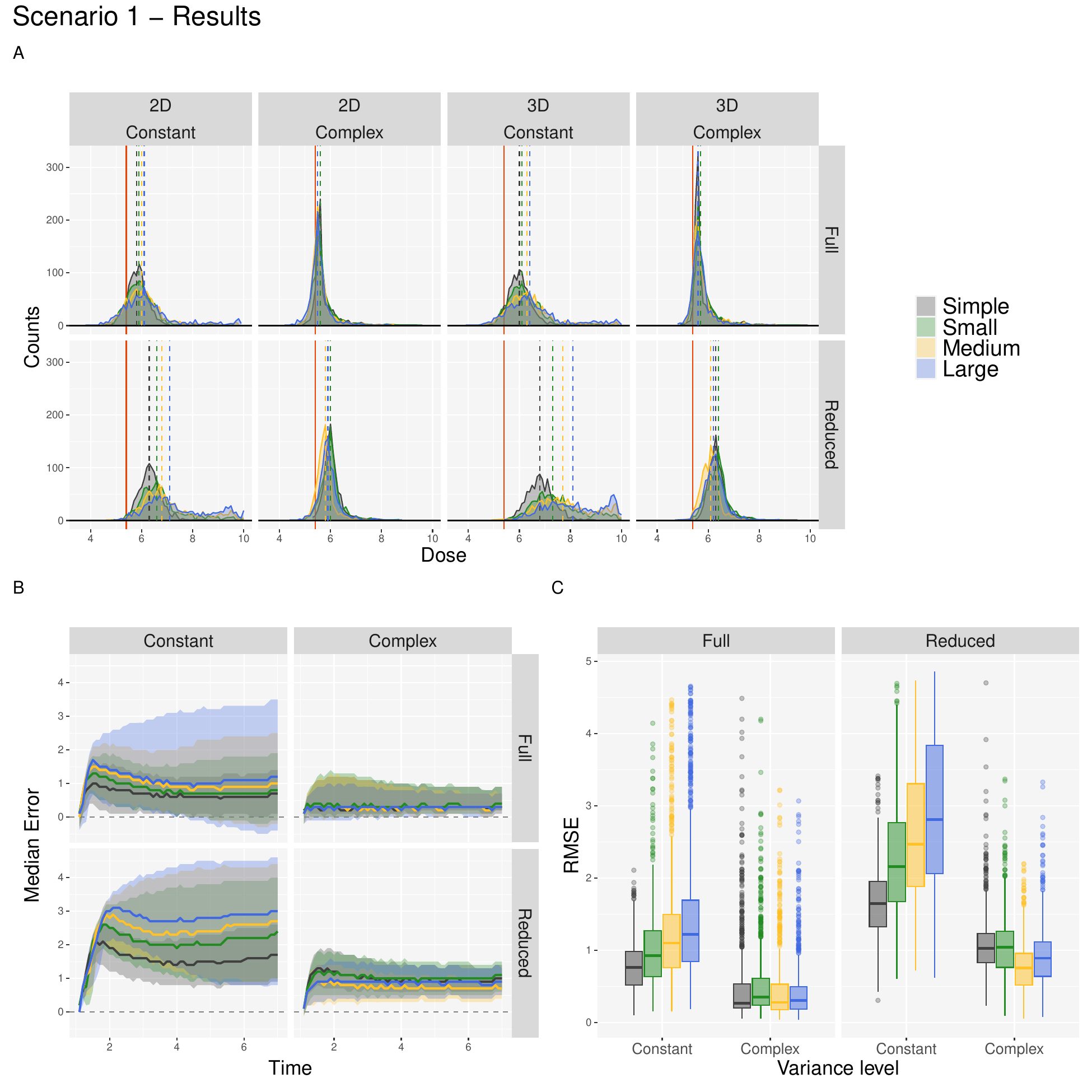}
    \caption{ (A) Counts of identified alerts at different dose levels for the two-dimensional approach. Each block contains the results of a model assumption regarding $\sigma$ (constant or complex) for the full or reduced design, respectively. The colors differentiate between the various simulation scenarios with simple, small, medium and large standard deviation. The red vertical line indicates the true alert. The dashed vertical lines show the respective medians. (B) Median error of the identified alert doses along the time-axis for the three-dimensional approach. Each block contains the results of a model assumption regarding $\sigma$ (constant or complex) for the full or reduced design, respectively. The solid line indicates the median. The areas indicate the $10\%$ to $90\%$ quantiles. (C) Boxplots of the RMSEs for the three-dimensional approach. 
    }
    \label{fig:resscen1}
\end{figure}

\begin{figure}[h!]
    \centering
    \includegraphics[width=0.7\linewidth]{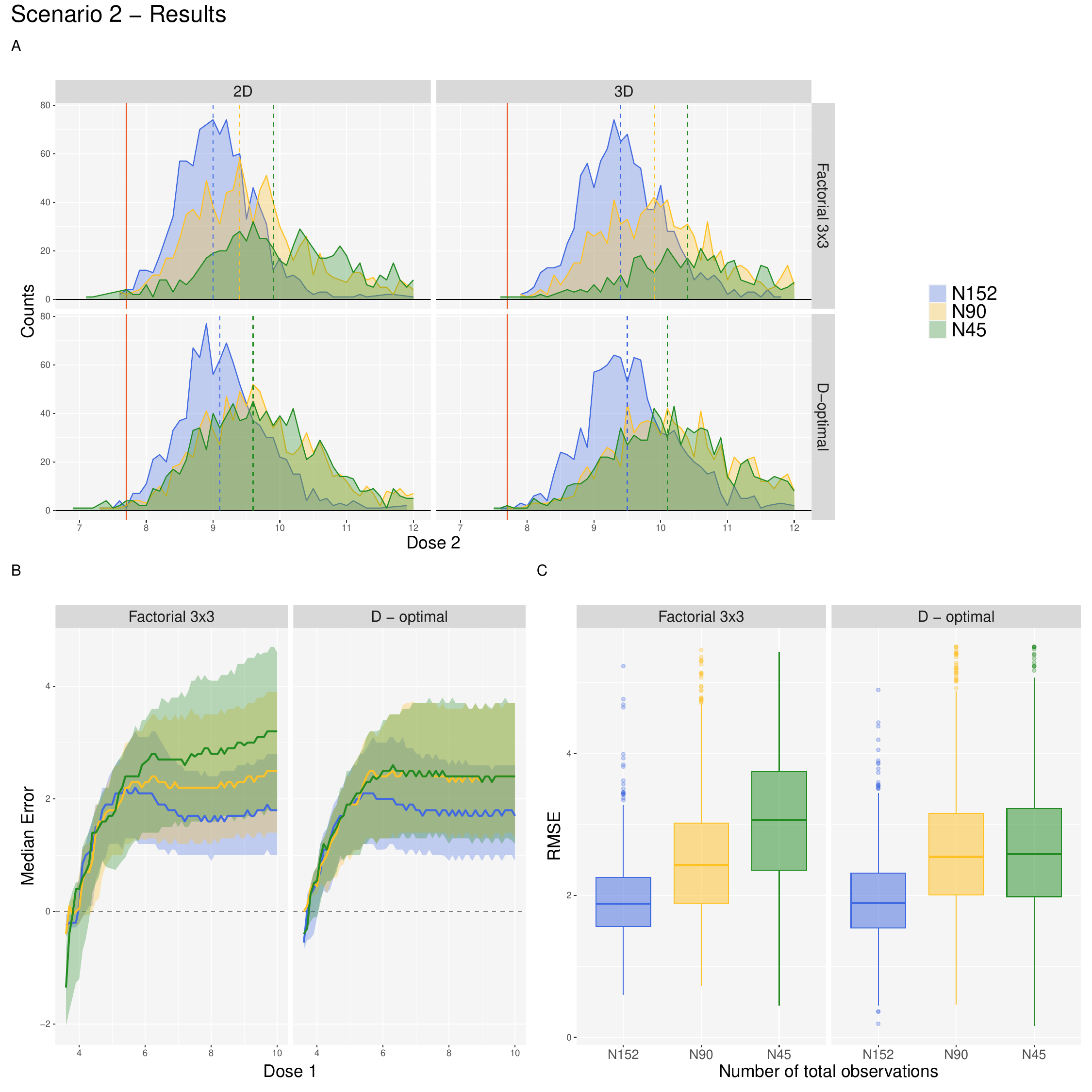}
    \caption{ (A) Counts of identified alerts at different levels of Dose 1 for the two-dimensional approach. Each block contains the results of the three different total numbers of observations.  The red vertical line indicates the true alert. The dashed vertical lines show the respective medians. (B) Median error of the identified alerts of Dose 2 along the $Dose \ 1$-axis for the three-dimensional approach. Each block contains the results of the three different total numbers of observations. The solid line indicates the median. The areas indicate the $10\%$ to $90\%$ quantiles. (C) Boxplots of the RMSEs for the three-dimensional approach. }
    \label{fig:resscen2}
\end{figure}

Figures \ref{fig:resscen1} (A) and \ref{fig:resscen2} (A) depict the number of the identified alerts per dose level (of Dose 2) and their respective medians for the different cases, compared to the true alert. The medians and the truth are indicated by vertical dashed lines and a solid line, respectively. The colors represent the simulated scenarios for the full (upper row) or reduced (lower row) design. The blocks show the results for either the two-dimensional or three-dimensional approach, assuming a constant or complex structure for the standard deviation. For Scenario 1, it is evident that the medians are closest to the truth for the full design with the two-dimensional approach, as expected. Furthermore, a decline in precision is observed as the standard deviation increases. These trends are visibly less pronounced for the complex assumption. Even in the simple case, the medians are closer to the truth. Additionally, the count distribution is visibly narrower. This clearly shows the benefits of the GAMLSS framework.

A similar trend is observed for the median and the number of observations in Scenario 2 with the factorial $3\times3$ design. However, the results for $n=90$ and $n=45$ are very similar for the D-optimal design. Additionally, the distribution appears slightly narrower in these cases.
\textcolor{black}{A general observation of the results for Scenario 2 reveals the presence of a bias, that is not as evident in the results of Scenario 1. 
As discussed in Section \ref{sec:model}, the assumption of a constant standard deviation might not properly reflect the complex structure of three-dimensional data and lead to a larger variation in simulated data sets. 
This effect occurs at two levels: Firstly, on the level of the simulated data sets themselves. Secondly, on both levels of the bootstrap procedure. Consequently, variability is amplified throughout the entire simulation setting.
This likely  impacts on the overall results and reinforces the method's tendency to underestimate the alert already observed for the simpler two-dimensional case in \citet{MS22}. This further supports the recommendation outlined in Section \ref{sec:model} to use GAMLSS for the modeling, in order to reduce the technical variability in favor of the inclusion of variability found in the data itself.  However, this was not possible to include in Scenario 2 since the parameters were extracted from an external source without a real data basis. 
To properly evaluate the procedure's performance 
and to do justice to the complex nature of three-dimensional data an additional simulation setting similar to the complex case in Scenario 1 for Scenario 2 would be necessary.}
\textcolor{black}{These findings are supported by the observation of a similar, albeit less pronounced, effect in the comparable sub-scenario of Scenario 1. This sub-scenario is the simple case in which the standard deviation is correctly assumed to be constant.}
The results for median and standard deviation are deferred to Tables B-E in the supplementary material.

\subsection{Results of the three-dimensional hypothesis}\label{sec:simres2d}
In this section, the results for the three-dimensional approach are discussed. 
Tables F-I in the supplementary material show the proportions of rejections of $H_0$ in $1,000$ simulation runs as well as the mean and standard deviation of the recall, precision, onset error and offset error. Similar trends in the number of rejections of $H_0$ are observed for three-dimensional hypotheses as for two-dimensional hypotheses. The mean recall for Scenario 1 is consistently high, with values exceeding $94.9\%$ for the full design and $86.3\%$ for the reduced design. In all the cases that were considered, and for all  the different assumptions, the standard deviation of the recall is observed to exceed $5\%$  once. The precision is even higher, almost $100\%$ with a standard deviation of $0.1\%$ or smaller. It can be observed that the mean offset error is consistently smaller than the onset error.  In fact, it is almost always $0$. This is possibly due to the offset equaling the largest time point observed in the study. Assuming a complex structure for the standard deviation results in a smaller mean onset error than assuming a constant standard deviation, except in a single case (Full, Simple, Normal), and the error is subject to smaller fluctuations between the cases overall. For the constant assumption, an increasing trend with larger simulated standard deviations can be observed. Both again show the benefit of the utilization of the GAMLSS framework.
However, the standard deviation of the onset error is slightly larger overall for the complex assumption in the full design. This is not the case for the reduced design, where the standard deviation for the complex assumption is smaller, except in the simple case. For Scenario 2, smaller means for recall and onset error are observed with the D-optimal design than with the factorial $3\times3$ design, while the respective standard deviations are similar in size. The results for precision and offset error are comparable to those for Scenario 1.

Figure \ref{fig:resscen1} (B) and Figure \ref{fig:resscen2} (B) depict the median error for the estimated alert dose (of Dose 2, respectively) and the $10\%$ and $90\%$ quantiles per time point (dose level of Dose 1, respectively). For Scenario 1, it is evident that the median error is smaller, and the quantiles are more proximate under the complex assumption. In Scenario 2, the discrepancy between the designs is less pronounced. 
However, the median error for the D-optimal design appears to be less significant for higher doses of Dose 1 in comparison to the factorial $3\times 3$ design. The results for Scenario 2 again show the bias discussed above for the two-dimensional results.

Figure \ref{fig:resscen1} (C) and Figure \ref{fig:resscen2} (C) depict boxplots of the RMSE for the various cases. For Scenario 1, the boxplots indicate smaller values overall for the complex assumption. 
The size of the boxplots for the constant assumption increases with the simulated standard deviation, which, in this case, can be interpreted as a degree of misspecification,  since the coefficients in \eqref{sigmaform} except for the intercept differ more from $0$. The size of the boxplots remains constant when assuming a complex structure, thereby underscoring the benefit of the GAMLSS framework. For Scenario 2, the absence of a trend is again observed in the results of D-optimal designs for the two smaller numbers of observations, which is clearly visible for the factorial $3\times 3$ design.

\section{Case Study}\label{sec:case}

In this section, an application of the proposed approach is presented. For illustration, the data set generated in \citet{Gu18} is used, in which a total of 30 compounds were  assessed in order to ascertain  the relevance of the exposure duration on cytotoxicity in primary human hepatocytes. Precisely, the goal was to determine whether additional information can be gained from  a prolonged  incubation period  beyond the usually used 1 to 2 days \citep{bsp12day1,bsp12day2}. As demonstrated by \citet{riddell1986importance},  substantial differences in the toxicity of compounds can be observed between 1 and 3 days. In the study of \citet{Gu18}, an additional third exposure period of 7 days was chosen. Based on statistical significance, the proposed approach 
can assist in determining whether a shorter time period would have sufficed in some cases. The following analysis focuses on the results generated for \textit{Aspirin} (\textit{ASP}).

For \textit{ASP}, cell viability data is available for 6 concentrations and exposure durations of 1,2 and 7 days.
The concentrations were  selected equidistant on a log scale with base $=\sqrt{10}$. The experimental design comprises 3 biological replicates (\textit{Donors}) and 4 or 8 measurements per combination of donor, exposure duration, and concentration. The data is normalized as described in \citet{DH22}.
In addition to the full data set, this section includes an analysis of an artificially reduced data set. This is used to demonstrate the approach in a setting with a smaller number of observations. The reduction is achieved by calculating the mean response of each donor for all combinations of exposure duration and concentration, and treating it as a single measurement.

Note that in the following, the terms \textit{time} and \textit{dose} are used instead of \textit{exposure duration} and \textit{concentration} to ensure consistency with the rest of this manuscript.

For cell viability, it can be assumed that following normalization, the response is $100\%$ at the control and tends to $0\%$ at higher concentrations. Hence, the simplified approach for the monotonically decreasing case as described in \eqref{hypothesesdecreasing} can be applied.
As a threshold value, $50\%$ cell viability was chosen. In this case, the corresponding alert is the value $ED_{50}$, which is frequently selected to evaluate the potency of a compound \citep{ec50}. For the two-dimensional analysis, the fixed time point was set at 4 days, which is one day longer than the longest time period examined in \citet{riddell1986importance} and recommended for many tests in \citet{GuidanceICCVAM}. 
For the parametric model fitting of the mean, the models proposed in \citet{DH22} (\textit{td2pLL}) and \citet{schuettler} (denoted \textit{schuettler} in the following) are selected as competitors, see the supplementary material for details on the parametrization. The following two  linear models for the standard deviation are considered: 
\begin{align*}
    \text{Constant:  }&  \log(\sigma) = \beta_{Intercept} \\
    \text{Complex: }& \log(\sigma) = \beta_{Intercept} +\beta_{Dose}Dose + \beta_{Dose^2}Dose^2 + \beta_{Time}Time + \beta_{Dose^2\cdot Time}Dose^2\cdot Time 
\end{align*}
Preliminary investigations indicated that the donor's influence is negligible and, consequently, has been excluded from consideration. 
 Note that the formula of the complex assumption was used in \citet{DH22} to generate a simulation study based on \citet{Gu18}. 
Also, note that the link function is applied in both cases as described in Section \ref{sec:meth:est}. By comparing the simple assumption with the more complex alternative, the benefits of the flexibility introduced by the proposed method can be evaluated. Figures D-I in the supplementary material depict the estimated linear models.
Model selection was conducted using the AIC \citep{FK22}.
However, given the small differences in AICs, the findings obtained from fitting the respective other parametric model are shown in the supplementary material for comparison. Overall, these results are very similar to those presented in the following.

\begin{figure}[h]
    \centering
    \includegraphics[width=0.7\linewidth]{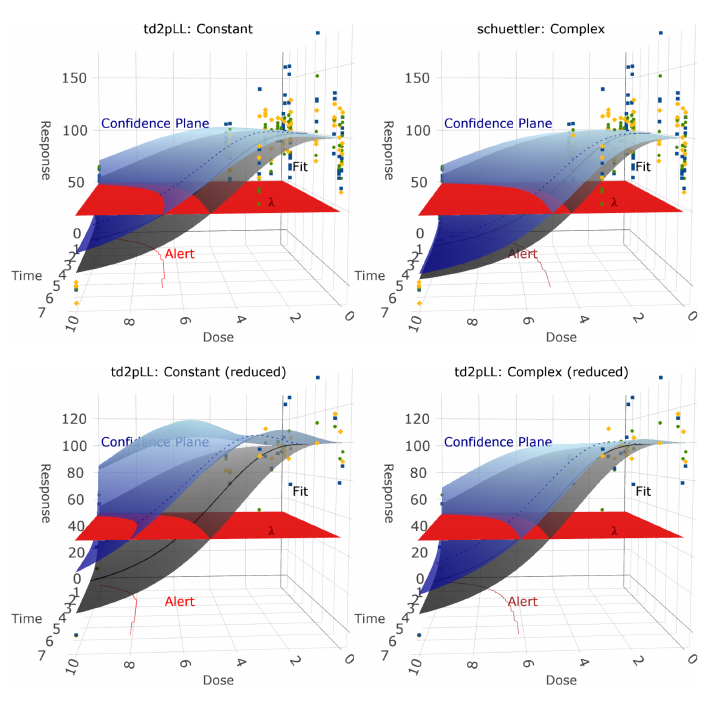}
            \caption{Visualization of the results obtained from the three-dimensional approach. The upper row displays the results for the full data set, the lower row the results for the artificially reduced data set. The dots represent the measured data points, which have been colored according to the donor. The black plane indicates the model fit. The blue plane represents the estimated confidence plane. The fixed time point $\tilde t=4$ is highlighted in both planes. The alert threshold $\lambda$ is displayed as a red hyperplane  that is orthogonal  to the $Response$-axis. The estimated alert relationship is projected onto the plane spanned by the dose-axis and the time-axis intersecting the $Response$-axis at the origin.}
                    \label{fig:CS3D}
\end{figure}

Figure \ref{fig:CS3D} displays the model fit  and the estimated confidence plane for each case. The fixed time point $\tilde t=4$ is highlighted. The red hyperplane indicates the chosen threshold value of $50\%$.  The significant time-dose-alert relationship identified is projected onto the hyperplane spanned by the dose-axis and the time-axis.
Overall, it can be observed that the estimated confidence plane  derived from the more complex model encloses the model fit more smoothly and identifies smaller alert doses for earlier time points. The former effect is especially visible for the reduced data set. The estimated confidence plane assuming the constant model demonstrates a discernible bulk for shorter time periods, which is not evident in the confidence plane estimated from the complex model. 

\begin{figure}[h]
    \centering
\includegraphics[width=0.7\linewidth]{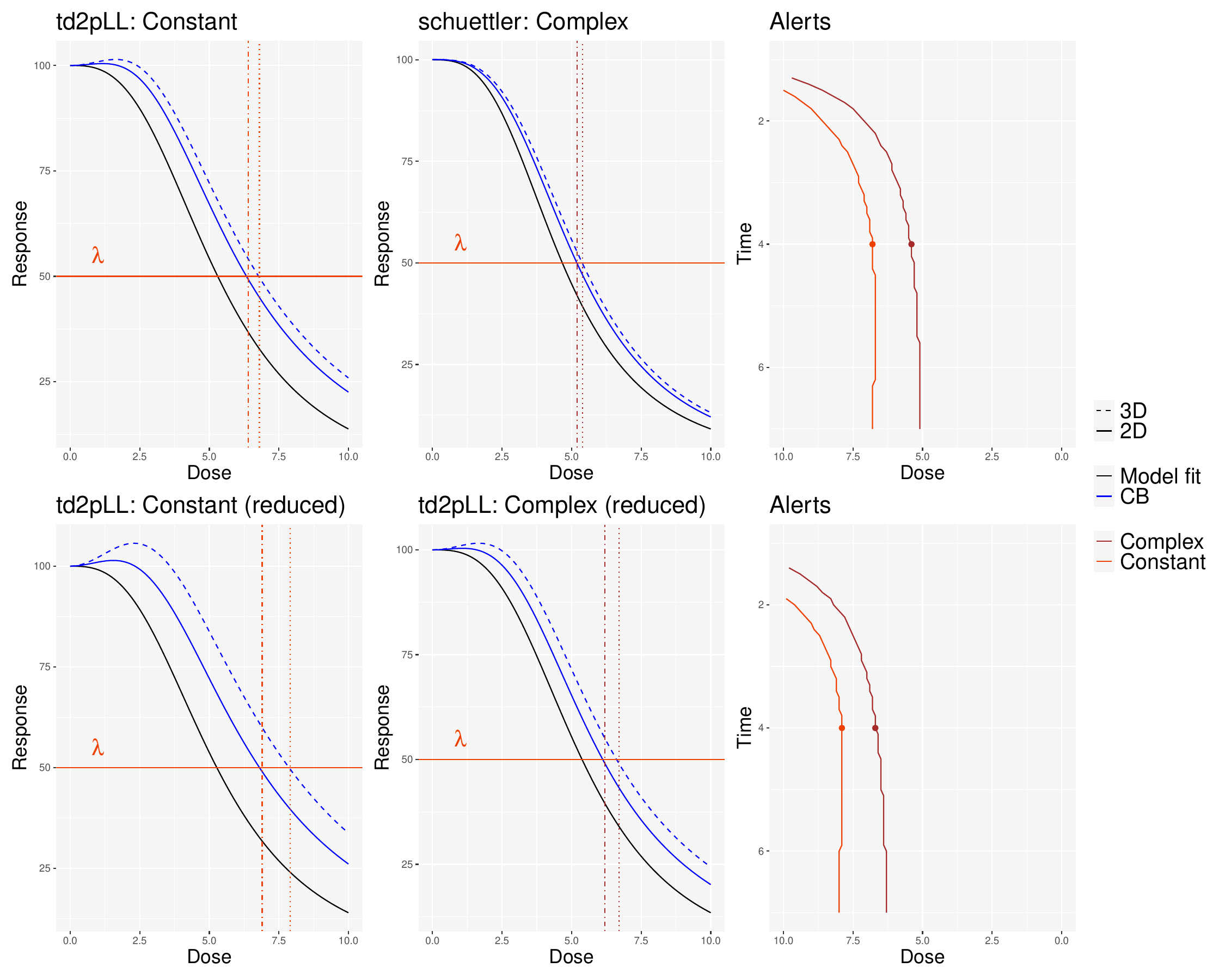}
  
            \caption{Visualization of the two-dimensional results at the fixed time point $\tilde t=4$. The upper row shows the results for the full data set, the lower row for the artificially reduced data set. On the left, the estimated  model fit  and the confidence bands (CB) are displayed. These have been estimated from the two-dimensional (solid) and three-dimensional (dashed) approaches. The red horizontal line indicates the threshold $\lambda=50\%$. The red and brown vertical lines highlight the respective identified alert dose. On the right, the estimated time-dose-alert relationship from the three-dimensional analysis is displayed. The dots indicate the alert for the fixed time point.}
              \label{fig:CS2D}
\end{figure}

On the left, Figure \ref{fig:CS2D} displays the two-dimensional results for $\tilde t=4$. As previously observed by means of the definition, the estimated confidence band projected from the three-dimensional approach is wider and identifies larger alert doses than the  confidence band derived from the two-dimensional approach. However, the difference is small. This indicates only a minor loss in accuracy of the three-dimensional approach on the two-dimensional level.

For the full data set, the identified alerts range from $6.4$ (2D) to $6.8$ (3D) for the simple model and from $5.2$ (2D) to $5.4$ (3D) for the complex model. The alerts identified for the artificially reduced data set are larger and extend over a wider interval, as was expected due to the uncertainty introduced by a smaller number of observations. For the constant model, the range is from  $6.9$ (2D) to $7.9$ (3D), whereas for the complex model it is from $6.2$ (2D) to $6.7$ (3D).
All identified alerts are collected in Tables J-K in the supplementary material.

On the right, Figure \ref{fig:CS2D} displays the two-dimensional time-dose-alert relationship, as identified by applying the three-dimensional approach. The difference between the constant model and the complex model is evident, with the doses identified for the complex case being smaller throughout the entire time span.
The shape of the identified alert-relationship is consistent across all cases.
For a brief period at the earliest time points, the maximum dose of 10 is too small to identify an alert. Subsequently, the identified alert decreases substantially up to approximately day 3. Thereafter, the decrease slows until the alert becomes almost constant around day 4.
Therefore, the findings suggest that a shorter exposure period might have been sufficient for \textit{ASP}.

\section{Discussion}\label{sec:dis}
In this paper, a parametric methodology for the identification of alerts within a complex, three-dimensional data structure is proposed. This is achieved by focusing on a fixed time point and applying a two-dimensional hypothesis test, or by applying a three-dimensional hypothesis test to identify a time-dose-alert relationship. Thereby, we provide a tool that enables statistically meaningful interpretation of not only single alert doses, but also of a two-dimensional relationship. 
As demonstrated by Scenario 1 in the simulation study, the integration of the GAMLSS framework in the bootstrap algorithm allows for a more precise analysis of complex standard deviation structures. The procedure can be applied to a broad range of statistical questions and a wide variety of models, as evidenced by the application of the  \textit{td2pLL}, \textit{schuettler} and \textit{Emax-Emax} models. As demonstrated by Scenario 2, in combination with an optimized design, the approach works stable even with a reduced number of observations. By introducing the fast approach, situations in which multiple runs are necessary and computation time is an issue is also considered.

Of course, this model has its limitations. As demonstrated by the simulation results in Section \ref{sec:sim} , there is a tendency to underestimate the true alert. The effect is relatively modest if the number of observations is not too small and the standard deviation structure is correctly specified. While the model appears relatively stable against misspecification of the standard deviation structure, the scope of the results presented in this manuscript may not be sufficient to provide a complete picture. Moreover, the approach is currently founded on model selection, particularly in regard to the parametric model selected for the mean.   A more flexible approach to model design, such as model averaging (see, for example, \citealp{hagemann2025overcoming}) or spline-based approaches (see \citealp{FK22}), might improve flexibility. Furthermore, the incorporation of random effects would be an interesting and important addition.
Also, the hypothesis test was conducted using a simultaneous confidence band/plane, that was estimated based on a procedure that extended the approach of \citet{MS22}. In principle, the hypothesis test may be conducted with any simultaneous confidence band/plane. A comparison of the various existing approaches may prove to be of interest.

In conclusion, we posit that our proposed method constitutes a valuable addition to the analysis of multidimensional dose-response data.  It enables a comprehensive analysis of both  two-dimensional and three-dimensional questions and
is applicable to a wide range of complex data structures. The extension to more than three continuous variables is a straightforward process, thereby enhancing its applicability.

\section*{Supplementary material}
The data presented in the case study are openly available in Archives of
Toxicology at 10.1007/s00204-018-2302-0 or included in the \texttt{R} package \textit{td2pLL}.
Supplementary material providing additional tables and figures can be found online. Corresponding R code, which can be used to reproduce the analysis of the case study and the simulation results, is available at
 \url{https://github.com/LuciaAmeis/Multivariate_alert_detection}.

\section*{Funding}

This work has been supported by the Research Training Group ''Biostatistical Methods for High-Dimensional Data in Toxicology'' (RTG 2624, P7) funded by the Deutsche Forschungsgemeinschaft (DFG, German Research Foundation - Project Number 427806116).


\begin{thebibliography}{}

\bibitem[Ameis and M{\"o}llenhoff, 2024]{ameis2024identification}
Ameis, L. and M{\"o}llenhoff, K. (2024).
\newblock Identification of changes in gene expression.
\newblock {\em arXiv preprint}.

\bibitem[Arbo et~al., 2016]{bsp12day1}
Arbo, M., Melega, S., St{\"o}ber, R., Schug, M., Rempel, E., Rahnenf{\"u}hrer,
  J., Godoy, P., Reif, R., Cadenas, C., {de Lourdes Bastos}, M., et~al. (2016).
\newblock Hepatotoxicity of piperazine designer drugs: up-regulation of key
  enzymes of cholesterol and lipid biosynthesis.
\newblock {\em Archives of Toxicology}, 90(12):3045--3060.

\bibitem[Bornkamp et~al., 2009]{bornkamp2009mcpmod}
Bornkamp, B., Pinheiro, J., and Bretz, F. (2009).
\newblock Mcpmod: An r package for the design and analysis of dose-finding
  studies.
\newblock {\em Journal of Statistical Software}, 29:1--23.

\bibitem[Box et~al., 2015]{box2015time}
Box, G., Jenkins, G., Reinsel, G., and Ljung, G. (2015).
\newblock {\em Time series analysis: forecasting and control}.
\newblock John Wiley \& Sons.

\bibitem[Bretz et~al., 2010]{bretz2010practical}
Bretz, F., Dette, H., and Pinheiro, J. (2010).
\newblock Practical considerations for optimal designs in clinical dose finding
  studies.
\newblock {\em Statistics in medicine}, 29(7-8):731--742.

\bibitem[{De Bastiani} et~al., 2026]{Gamlssside}
{De Bastiani}, F., Heller, G., Kneib, T., Mayr, A., Rigby, R., Stasinopoulos,
  M., et~al. (2026).
\newblock {\em {GAMLSS}}.
\newblock Accessed February 05, 2026.

\bibitem[Delignette-Muller et~al., 2011]{DF11}
Delignette-Muller, M.-L., Forfait, C., Billoir, E., and Charles, S. (2011).
\newblock {A new perspective on the Dunnett procedure: Filling the gap between
  NOEC/LOEC and ECx concepts}.
\newblock {\em Environmental Toxicology and Chemistry}, 30(12):2888--2891.

\bibitem[Demidenko, 2013]{demidenko2013mixed}
Demidenko, E. (2013).
\newblock {\em Mixed models: theory and applications with R}.
\newblock John Wiley \& Sons.

\bibitem[Duda et~al., 2023]{DD23}
Duda, J., Drenda, C., K\"astel, H., Rahnenf\"uhrer, J., and Kappenberg, F.
  (2023).
\newblock Benefit of using interaction effects for the analysis of
  high-dimensional time-response or dose-response data for two-group
  comparisons.
\newblock {\em Scientific Reports}, 13(1):20804.

\bibitem[Duda et~al., 2022]{DH22}
Duda, J., Hengstler, J., and Rahnenf\"uhrer, J. (2022).
\newblock td2pll: An intuitive time-dose-response model for cytotoxicity data
  with varying exposure durations.
\newblock {\em Computational Toxicology}, 23:100234.

\bibitem[Efron and Tibshirani, 1994]{ET94}
Efron, B. and Tibshirani, R. (1994).
\newblock {\em {An Introduction to the Bootstrap}}.
\newblock CRC Press, 1st ed. edition.

\bibitem[Fahrmeir et~al., 2022]{FK22}
Fahrmeir, L., Kneib, T., Lang, S., and Marx, B. (2022).
\newblock {\em Regression: Models, Methods and Applications}.
\newblock Springer Berlin, Heidelberg, 2nd edition.

\bibitem[Fawcett, 2006]{fawcett2006introduction}
Fawcett, T. (2006).
\newblock An introduction to roc analysis.
\newblock {\em Pattern recognition letters}, 27(8):861--874.

\bibitem[Ga{\l}ecki and Burzykowski, 2012]{galecki2012linear}
Ga{\l}ecki, A. and Burzykowski, T. (2012).
\newblock Linear mixed-effects model.
\newblock In {\em Linear mixed-effects models using R: a step-by-step
  approach}, pages 245--273. Springer.

\bibitem[Ghallab et~al., 2016]{bsp12day2}
Ghallab, A., Celli{\`e}re, G., Henkel, S., Driesch, D., Hoehme, S., Hofmann,
  U., Zellmer, S., Godoy, P., Sachinidis, A., Blaszkewicz, M., et~al. (2016).
\newblock Model-guided identification of a therapeutic strategy to reduce
  hyperammonemia in liver diseases.
\newblock {\em Journal of hepatology}, 64(4):860--871.

\bibitem[Gu et~al., 2018]{Gu18}
Gu, X., Albrecht, W., Edlund, K., Kappenberg, F., Rahnenf{\"u}hrer, J., Leist,
  M., Moritz, W., Godoy, P., Cadenas, C., Marchan, R., et~al. (2018).
\newblock Relevance of the incubation period in cytotoxicity testing with
  primary human hepatocytes.
\newblock {\em Archives of toxicology}, 92(12):3505--3515.

\bibitem[Hagemann and M{\"o}llenhoff, 2025]{hagemann2025overcoming}
Hagemann, N. and M{\"o}llenhoff, K. (2025).
\newblock Overcoming model uncertainty—how equivalence tests can benefit from
  model averaging.
\newblock {\em Statistics in Medicine}, 44(6):e10309.

\bibitem[Hothorn, 2014]{Hot14}
Hothorn, L. (2014).
\newblock Statistical evaluation of toxicological bioassays - a review.
\newblock {\em Toxicology Research}, 3(6):418--432.

\bibitem[Huusari et~al., 2025]{huusari2025predicting}
Huusari, R., Wang, T., Szedmak, S., Aittokallio, T., and Rousu, J. (2025).
\newblock Predicting drug combination response surfaces.
\newblock {\em npj Drug Discovery}, 2(1):2.

\bibitem[{ICCVAM}, 2001]{GuidanceICCVAM}
{ICCVAM} (2001).
\newblock {\em Guidance document on using in vitro data to estimate in vivo
  starting doses for acute toxicity}.
\newblock Accessed March 18, 2026.

\bibitem[Jensen et~al., 2019]{JK19}
Jensen, S., Kluxen, F., and Ritz, C. (2019).
\newblock A review of recent advances in benchmark dose methodology.
\newblock {\em Risk Analysis}, 39(10):2295--2315.

\bibitem[Kappenberg et~al., 2023]{KD23}
Kappenberg, F., Duda, J., Sch{\"u}rmeyer, L., G{\"u}l, O., Brecklinghaus, T.,
  Hengstler, J., Schorning, K., and Rahnenf{\"u}hrer, J. (2023).
\newblock Guidance for statistical design and analysis of toxicological
  dose--response experiments, based on a comprehensive literature review.
\newblock {\em Archives of Toxicology}, 97(10):2741--2761.

\bibitem[Kappenberg et~al., 2021]{KG21}
Kappenberg, F., Grinberg, M., Jiang, X., Kopp-Schneider, A., Hengstler, J., and
  Rahnenf\"uhrer, J. (2021).
\newblock Comparison of observation-based and model-based identification of
  alert concentrations from concentration-expression data.
\newblock {\em Bioinformatics}, 37(14):1990 -- 1996.

\bibitem[Lange and Schmidli, 2015]{lange2015analysis}
Lange, M. and Schmidli, H. (2015).
\newblock Analysis of clinical trials with biologics using dose--time-response
  models.
\newblock {\em Statistics in Medicine}, 34(22):3017--3028.

\bibitem[Lee, 2010]{lee2010drug}
Lee, S. (2010).
\newblock Drug interaction: focusing on response surface models.
\newblock {\em Korean journal of anesthesiology}, 58(5):421--434.

\bibitem[Loewe, 1953]{loewe1953problem}
Loewe, S. (1953).
\newblock The problem of synergism and antagonism of combined drugs.
\newblock {\em Arzneimittel-forschung}, 3(6):285--290.

\bibitem[Manning, 2008]{manning2008introduction}
Manning, C. (2008).
\newblock {\em Introduction to information retrieval}.
\newblock Syngress Publishing,.

\bibitem[Martin-Betancor et~al., 2015]{martin2015defining}
Martin-Betancor, K., Ritz, C., Fern{\'a}ndez-Pi{\~n}as, F., Legan{\'e}s, F.,
  and Rodea-Palomares, I. (2015).
\newblock Defining an additivity framework for mixture research in inducible
  whole-cell biosensors.
\newblock {\em Scientific Reports}, 5(1):17200.

\bibitem[Mokhtari et~al., 2017]{mokhtari2017combination}
Mokhtari, R., Homayouni, T., Baluch, N., Morgatskaya, E., Kumar, S., Das, B.,
  and Yeger, H. (2017).
\newblock Combination therapy in combating cancer.
\newblock {\em Oncotarget}, 8(23):38022.

\bibitem[M\"ollenhoff et~al., 2022]{MS22}
M\"ollenhoff, K., Schorning, K., and Kappenberg, F. (2022).
\newblock Identifying alert concentrations using a model-based bootstrap
  approach.
\newblock {\em Biometrics}, 79(3):2076--2088.

\bibitem[Riddell et~al., 1986]{riddell1986importance}
Riddell, R., Panacer, D., Wilde, S., Clothier, R., and Balls, M. (1986).
\newblock The importance of exposure period and cell type in in vitro
  cytotoxicity tests.
\newblock {\em Alternatives to Laboratory Animals}, 14(2):86--92.

\bibitem[Rigby and Stasinopoulos, 2005]{rigby2005generalized}
Rigby, R. and Stasinopoulos, D. (2005).
\newblock Generalized additive models for location, scale and shape.
\newblock {\em Journal of the Royal Statistical Society Series C: Applied
  Statistics}, 54(3):507--554.

\bibitem[Sch{\"u}rmeyer et~al., 2025]{Sch25}
Sch{\"u}rmeyer, L., Sandig, L., Hezler, L., Igl, B.-W., and Schorning, K.
  (2025).
\newblock Optimal designs for identifying effective doses in drug combination
  studies.
\newblock {\em arXiv preprint}.

\bibitem[Sch\"uttler et~al., 2019]{schuettler}
Sch\"uttler, A., Altenburger, R., Ammar, M., Bader-Blukott, M., Jakobs, G.,
  Knapp, J., Krüger, J., Reiche, K., Wu, G., and Busch, W. (2019).
\newblock Map and model—moving from observation to prediction in
  toxicogenomics.
\newblock {\em GigaScience}, 8(6):giz057.

\bibitem[Sebaugh, 2011]{ec50}
Sebaugh, J.~L. (2011).
\newblock Guidelines for accurate ec50/ic50 estimation.
\newblock {\em Pharmaceutical Statistics}, 10(2):128--134.

\bibitem[Stasinopoulos and Rigby, 2008]{GAMLSSRpaper}
Stasinopoulos, D. and Rigby, R. (2008).
\newblock Generalized additive models for location scale and shape (gamlss) in
  r.
\newblock {\em Journal of Statistical Software}, 23:1--46.

\bibitem[Stasinopoulos et~al., 2017]{GAMLSSbook}
Stasinopoulos, M., Rigby, R., Heller, G., Voudouris, V., and {De Bastiani}, F.
  (2017).
\newblock {\em Flexible Regression and Smoothing: Using GAMLSS in R}.
\newblock Chapman and Hall/CRC.

\bibitem[Zhou et~al., 2025]{zhou2025combination}
Zhou, Y., Sloan, A., Menon, S., and Wang, L. (2025).
\newblock Combination mcp-mod for two-drug combination dose-ranging studies.
\newblock {\em Journal of Biopharmaceutical Statistics}, 35(2):257--270.

\end{thebibliography}

\end{document}